\documentstyle[12pt,aaspp4]{article}

\slugcomment{To appear in The Astrophysical Journal}

\begin{document}

\title{ Synthesis Imaging of Dense Gas in Nearby Galaxies}

\author{Tamara T. Helfer\altaffilmark{1} and Leo Blitz\altaffilmark{1}}
\affil{Department of Astronomy, University of Maryland,
    College Park, MD 20742}

\altaffiltext{1}{thelfer,blitz@astro.berkeley.edu, current address 
Radio Astronomy Lab, 601 Campbell Hall, UC Berkeley, Berkeley CA 94720}

\begin{abstract}

We present images of the HCN J = 1-0 emission from 
five nearby spiral galaxies made with the Berkeley-Illinois-Maryland 
Association interferometer.  The HCN observations comprise the first 
high-resolution ($\theta$ $\sim$ 5\arcsec -- 10\arcsec) survey of dense 
molecular gas from a sample of normal galaxies, rather than galaxies with 
prolific starburst or nuclear activities.  The images show compact 
structure, demonstrating that the dense gas emission is largely confined to the
central kiloparsec of the sources.  To within the uncertainties,
70 - 100\% of the single-dish flux is recovered for
each source; this implies that there is not a significant contribution
to the HCN flux from low-level emission in the disks of the galaxies.
In one of the galaxies, NGC 6946, the ratio of HCN to CO integrated
intensities ranges from 0.05--0.2 within the extent of the HCN emission
($r = 150$ pc), with an average value of 0.11 $\pm$ 0.01 over the whole
region; the range and average values of the ratios in NGC 6946 are
very similar to what is observed in the central $r = 250$ pc of the
Milky Way.  A comparison with single-dish observations allows us to place 
an upper limit of 0.01 on the ratio of integrated intensities in the 
region $150 < r < 800$ pc in NGC 6946.  In NGC 6946, NGC 1068 and 
the Milky Way, the ratio I$_{\rm HCN}$/I$_{\rm CO}$ is 5 to 10 times 
higher in the bulge regions than in their disks; this suggests that the 
physical conditions in their bulges and disks are very different.  
Furthermore, the presence of dense gas on size scales of $\sim$ 500 pc
in the centers of these nearby galaxies and the Milky Way suggests that 
the internal pressure is at least 10$^7$ cm$^{-3}$~K in their centers; 
this is some three orders of magnitude greater than the pressure in the 
local interstellar medium in the Milky Way, and it is two orders of magnitude 
greater than the pressure from the self-gravity of a solar neighborhood 
giant molecular cloud. 
In NGC 4826 and M51, as in the Milky Way and NGC 1068, there is a linear 
offset of $\sim$ 100 pc between the dense gas distribution and the peak 
of the radio continuum emission.  We did not detect HCN towards three 
additional spiral galaxies.  

\end{abstract}


\section{ Introduction }

Single dish millimeter observations of HCN and CS emission from nearby 
galaxies show that most spiral galaxies, not just starburst galaxies, 
have an appreciable amount of dense ($\sim$ 10$^5$ cm$^{-3}$) gas in their 
bulges (Helfer \& Blitz 1993 and references therein).  These observations 
are in good agreement with what is seen in the inner $r \sim 250$ parsecs of 
the Milky Way, where strong, diffuse emission from CS (Bally et al. 1987) 
and HCN (Jackson et al. 1996) is observed despite the 
moderate star formation rates there.  (HCN and CS trace gas densities of
$\ga$ 10$^5$ cm$^{-3}$ in galaxies; this is two orders of magnitude higher
than the density required to excite CO.)  The ubiquity of large-scale emission 
from HCN and CS over hundreds of parsecs in the centers of galaxies is 
surprising when compared with the known properties of giant molecular clouds 
(GMCs) in the solar neighborhood, which contain few and relatively 
small ($\la$ pc-scale) clumps of dense gas --- and these only where stars 
are actively forming or where stars have formed very recently (i.e. where 
there is a local source of pressure, not common to the GMC as a whole).

In galaxies, single-dish millimeter wavelength beams typically cover tens to 
hundreds of GMC-sized diameters.  As part of our program to measure 
the degree to which different galactic environments affect 
the properties of molecular clouds, we began a program with the 
Berkeley-Illinois-Maryland Association (BIMA) interferometer to 
survey eight nearby galaxies in HCN emission in order to measure 
the distribution and amount of dense gas on size scales of individual 
GMCs or small associations of GMCs.  We consider these galaxies to be
``normal'' since they do not have prolific circumnuclear starbursts or 
active galactic nuclei; however, like most spiral galaxies, these galaxies 
have low-level nuclear line emission (Ho, Filippenko \& Sargent 1993).  
In this paper, we present the results of the BIMA survey.   We also present 
a comparison of the HCN emission from one of the sources, NGC 6946, with 
its CO J = 1-0 emission (Regan \& Vogel 1995).  A comparison with the CO 
emission from the other sources will be the topic of a future paper.

\section{ Observations }

\subsection{ BIMA Observations }

The sources and their coordinates are listed in Table 1.
The sources were selected from the single-dish survey of Helfer 
\& Blitz (1993) as galaxies with strong CO emission that were
also detected in HCN with a single pointing using the NRAO 12 m telescope. 
The data were collected with the BIMA interferometer (Welch et 
al. 1996), which then comprised 6 antennas,
between 1994 February 15 and 1995 May 02;  the observations
included data from up to three array configurations.  
For each source, the receivers were tuned to the redshifted frequency of 
the HCN J = 1-0 transition ($\nu$$_{\rm o}$ = 88.61 GHz).   The data
were processed using the MIRIAD package (Sault, Teuben, \& Wright
1995).  The time variations of the amplitude and phase gains were 
calibrated using short observations of nearby quasars every $\sim$ 30 
minutes;  a planet or strong quasar was observed to set the absolute flux 
scale as well as to calibrate the spectral dependence of the gains across the
IF passband.  The digital correlator was configured to achieve a
maximum spectral resolution of 1.56 MHz (5.3 km s$^{-1}$).  For some of 
the observations, there was an intermittent phase lock on one of the
antenna receivers that was not discovered until after the observations; in 
these cases all baselines involving that antenna were eliminated from further
data reduction.  

For each source, the calibrated data were smoothed to 21.1 km s$^{-1}$
resolution, then gridded and Fourier transformed using natural weighting.
The visibilities were also weighted by the inverse of the noise
variance, which was determined from the system temperatures and from
the gains of the individual antennas.  We cleaned the maps using the
standard H\"ogbom algorithm.   Maps of integrated intensity were made
by summing those channels with emission after clipping the channel maps
at a 1 $\sigma$ level.  The beam sizes and noise levels of the
final maps are listed in Table 2.  The noise levels are somewhat
underestimated relative to the formal uncertainties, since they
were determined from the clipped moment maps.  The absolute flux
calibration in all maps is probably accurate to $\pm$ 30\%.

\subsection{ Single-dish Observations of CO in NGC 6946}

We have also used BIMA to image CO in the centers of three of the galaxies 
(NGC 3628, NGC 4826, and M83) with detected HCN emission.   A fourth galaxy, 
NGC 6946, has been imaged in CO using BIMA by Regan \& Vogel (1995).  
The CO emission, unlike the HCN emission (see below), is
significantly ``resolved out''
in the interferometric images of these galaxies;  it is therefore necessary 
to fill in the zero-spacing flux with a single-dish telescope.   
We are currently still making these measurements for the first three
sources; we therefore defer further discussion of the CO results for
these galaxies to a future paper.

In the case of NGC 6946, we measured the CO short-spacing flux with the
NRAO 12 m telescope\footnote{The National Radio Astronomy Observatory 
is operated by Associated Universities, Inc., under cooperative agreement
with the National Science Foundation.} and combined these data with Regan
\& Vogel's BIMA map.  We observed on 95 June 20--21 in the newly implemented 
``on-the-fly'' (OTF) scheme at the NRAO 12 m (Emerson et al., in preparation) 
and covered a region roughly 5\arcmin\ on a side.   We observed orthogonal 
polarizations using two 256 channel filterbanks, each with a spectral 
resolution of 2 MHz per channel. The data were gridded and a linear baseline 
removed from the resulting data cubes in AIPS; the data were then transferred 
to the MIRIAD package for further processing.  Details of the single dish 
and interferometric data combination were very similar to those described in
Helfer \& Blitz (1995) for the case of NGC 1068.

\section{ Results }

We detected HCN in five of the eight galaxies:  NGC 3628,
NGC 4826, NGC 5194 (M51), NGC 5236 (M83), and NGC 6946; their images 
are presented in 
Figure \ref{hcnmaps} along with optical images from the Digitized Sky 
Survey.\footnote{Based on photographic data of the National Geographic
Society -- Palomar Observatory Sky Survey (NGS-POSS) obtained using the
Oschin Telescope on Palomar Mountain.  The NGS-POSS was funded by a grant
from the National Geographic Society to the California Institute of
Technology.  The plates were processed into the present compressed digital
form with their permission.  The Digitized Sky Survey was produced at the
Space Telescope Science Institute under US Government grant NAG W-2166.}
The spectra from the positions of peak HCN emission in each of the detected 
sources are shown in Figure \ref{spectra}.  

In each case, the HCN images show compact structure, with the detected
emission confined to the central $\sim$ 500 pc diameter; the 
5\arcsec--10\arcsec\ FWHM synthesized beam sizes correspond to linear
resolutions of 125--200 pc at the distances of the galaxies.  While the
HCN emission is compact, it appears resolved by the interferometer in 
each map.  In order to investigate the effect of the lack of $uv$ sampling 
at small spatial visibilities, we compared the integrated intensities 
measured with the interferometer to the single-dish fluxes of Helfer \& 
Blitz (1993).  Table 3 lists the integrated intensities measured at BIMA 
along with those measured at the NRAO 12 m and the fraction of the single-dish
flux recovered by the interferometer.  With the possible exception
of M51 (see $\S$ 3.1), the interferometer appears to recover all the 
single-dish flux measured by Helfer
\& Blitz.  This means that the maps shown in Figure \ref{hcnmaps} are 
a reliable representation of the distribution of the HCN emission, constrained
of course by the usual limitation of the signal to noise ratios.
Allowing for low-level, more diffuse emission detected at $< 2~\sigma$,
we conclude that the HCN emission is confined to the central kiloparsec 
of the galaxies (it is perhaps slightly more extended in M51).  The 
spatial extent of the HCN emission is in good agreement with what is seen 
in the Milky Way (Jackson et al. 1996) and in NGC 1068 (Helfer \& Blitz 1995).

\subsection{ Individual Sources }

The galaxies in this survey were originally selected from a list of 
the brightest extragalactic CO emitters (see Helfer \& Blitz 1993).
The following are brief descriptions of the individual sources.

NGC 3628 -- This galaxy is a member of the Leo triplet (along with
NGC 3623 and NGC 3627).  It is a nearly edge-on Sbc galaxy with a
prominent and irregular dust lane.   Its nuclear region contains
a modest starburst (e.g. Condon et al. 1982, Braine \& Combes 1992).  
The CO from NGC 3628 (Young, Tacconi, \& Scoville 1983; Boiss\'e, 
Casoli, \& Combes 1987; and Israel, Baas, \& Maloney 1990) is strongly 
peaked in the inner kiloparsec and has a similar extent and mass as 
that in the center of the Milky Way (Boiss\'e et al. 1987). 
The HCN in NGC 3628 is resolved and appears elongated in the east-west
direction.  There is emission at the 2 $\sigma$ level to the northwest 
of the central source that may be associated with gas further along the 
major axis.    The HCN, CO and radio continuum centers of NGC 3628
all peak some 21\arcsec\ to the southeast of the optical nucleus;
however, the optical nucleus is heavily obscured and its position is 
highly uncertain (Boiss\'e et al. 1987).  

NGC 4826 -- This Sab galaxy has been called variously the ``Black Eye,''
the ``Evil Eye,'' or somewhat more optimistically (Rubin 1994) the 
``Sleeping Beauty''
galaxy for its conspicuous dust lane (see Sandage 1961; see also 
the cover of {\it The Astronomical Journal}, 1994, 107, 1).  NGC 4826 has 
gained notoriety recently for the discovery that the inner kiloparsec-scale 
disk is counterrotating with respect to the larger scale rotation of the 
galaxy (Braun, Walterbos, \& Kennicutt 1992; Rubin 1994; Braun et al. 1994).   
The CO emission (Casoli \& Gerin 1993) is confined to the inner $r \sim 
1\arcmin$. The HCN emission appears symmetric and centrally peaked in NGC 4826.

NGC 5194 (M51) -- M51 is the prototypical grand design spiral and 
is one of the best studied galaxies in CO emission.  Interferometric
CO images of the nuclear region of M51 (e.g. Lo et al. 1987; Rand \& Kulkarni 
1990; and Adler et al. 1992; the last includes zero-spacing flux) show
a notable lack of a single central concentration of CO, despite the strong 
molecular emission associated with the spiral arms in the nuclear region.  
M51 is the only galaxy of the five detected in HCN at BIMA that does not 
have a strong central concentration of CO.    In contrast to the CO
emission, the HCN emission does appear to be centrally concentrated, 
though there is a significant contribution 
to the total flux from low-level ($< 2~\sigma$) emission. We note that
Kohno et al. (1996) mapped the central concentration as well as more 
extended structure in their HCN map of M51 observed with the Nobeyama 
Millimeter Array.  The linear extent of the low-level ($< 2~\sigma$)
HCN emission in the BIMA map is somewhat larger than those 
of the other four galaxies 
studied here, and the interferometer may have ``resolved out'' a nonnegligible
contribution to the total flux in this source (Table 3).  For M51,
the flux of structures larger than $\sim$ 18\arcsec\  is attenuated
by $\ga$~50\%.

NGC 5236 (M83) --  This well-studied source is a grand design, Sc/SBb 
spiral galaxy with strong circumnuclear star formation within the central 
few hundred pc (e.g. Gallais et al. 1991).  The HCN emission is strongest 
at the position of the peak of the radio continuum emission (Condon et al.
1982; Turner \& Ho 1994);  the condensations to the east and to the
south of the strongest HCN emission are also apparent in the radio
continuum.  The synthesized beam of the interferometer is quite elongated 
because of the foreshortening of the north-south baselines toward this low
declination source. The north-south elongation of the HCN emission is
therefore almost certainly an artifact of the observations.  

NGC 6946 --  This late-type, grand design spiral galaxy contains
a moderate starburst (Turner \& Ho 1983).   Within the inner
1.5 kpc diameter, the CO in this galaxy has a non-axisymmetric, 
north-south distribution that has been interpreted as a bar (Regan \&
Vogel 1995 and references therein).  However, Regan \& Vogel (1995) 
combined new CO and K-band observations and showed that the CO traces 
gas on the trailing side of spiral arms;  their observations are consistent
with what is expected for the gas and stellar response to a spiral
density wave rather than a bar.  The CO peaks up strongly in the inner 
300 pc diameter of NGC 6946. The HCN emission is detected in this region 
and is resolved and slightly extended in an east-west direction,
with an additional elongation towards the northwest (there is a similar
northwest extension in the CO map of Regan \& Vogel 1995).  We discuss this 
source more fully in the following section.

\subsection{ $\bf\rm I_{HCN}/I_{CO}$ in NGC 6946}

The ratio of the 3 mm integrated intensities, $\rm I_{HCN}/I_{CO}$,
may be used as a qualitative measure of the molecular gas density
(e.g. Helfer \& Blitz 1996).
In order to determine any line ratio from interferometric measurements, 
one must first take into account the possibility that the flux
measured with the interferometer is missing a significant contribution
from large-scale structures in the maps.  In NGC 6946, the interferometer 
recovers all the single-dish HCN flux to within the errors of the
measurement (Table 3).  For the CO, the Regan \& Vogel (1995) BIMA map
recovered about half the single-dish flux; we therefore modeled the 
short spatial frequency visibilities from the NRAO 12 m data ($\S$ 2.2) 
and combined these with the BIMA CO map.   The CO distribution and
flux in the resulting map did not change appreciably interior to
r = 15\arcsec; at larger radii, the most dramatic flux increases
were distributed over radii from 20\arcsec--50\arcsec, though the
shape of the structures remained about the same.  With the fully-sampled
CO map, we can now make a legitimate comparison of the HCN and CO intensities 
for this source.

To determine the ratio $\rm I_{HCN}/I_{CO}$ in NGC 6946, we convolved 
the CO map to match the resolution of the HCN image, and we
converted both intensities to a main beam brightness scale ($\rm \int 
T_{MB}$ $\Delta v$); the ratio was computed only for regions with
detected HCN emission (I$_{\rm HCN} > 2.5~\sigma_{\rm mom}$, or
r $\la$ 15\arcsec).  The resulting ratio map of I$_{\rm HCN}$/I$_{\rm CO}$
is presented in Figure \ref{n6946rat}.    
Although the emissions from CO and HCN 
both peak at the center of NGC 6946, the distribution of the HCN/CO ratio is
saddle-shaped, with the highest values to the east and west of the
nucleus by about 7\arcsec\ (175 pc) and the lowest values to
the northwest and southeast of the nucleus by about 5\arcsec\ (125 pc).
At first, it seems surprising that the ratio does not rise monotonically
to the central position.  However, at the small size scales resolved by the
interferometer ($r \approx 70$ pc), the characteristics of individual molecular
clouds start to dominate the distribution, rather than the integrated effects
of dozens of GMCs.  It may be that the small-scale ratio is dominated by
local effects from individual clouds; this effect is seen within the central 
few hundred pc of the Milky Way, where the I$_{\rm CS}$/I$_{\rm CO}$ ratio 
(Helfer \& Blitz 1993) and the I$_{\rm HCN}$/I$_{\rm CO}$ ratio (Jackson 
et al. 1996) look very clumpy and irregularly distributed.  In the 
Milky Way, features like the Sgr A and B GMCs are characterized by
relatively high values in the ratio maps.

The average ratio over the 
extent of the HCN emission ($\sim 12\arcsec$, or 300 pc diameter) 
in NGC 6946 is $0.11 \pm 0.01$, with a peak value of 0.19 
and a minimum of 0.049.  The range of
I$_{\rm HCN}$/I$_{\rm CO}$ in NGC 6946 is very similar to what is seen
in the Milky Way by Jackson et al. (1996);  on small scales within the
inner few degrees of the Milky Way, I$_{\rm HCN}$/I$_{\rm CO}$ ranges 
from 0.04 to 0.12, and the average over the extent of the HCN
emission, or r $\approx$ 300 pc, is 0.08 (see below).

We can compare the ratio we measure in the central 300 pc of NGC 6946
with those measured with larger apertures: the single dish ratios of 
I$_{\rm HCN}$/I$_{\rm CO}$ in NGC 6946 are 0.063 $\pm$ 0.007
at a resolution of 24\arcsec\ (600 pc) (Nguyen-Q-Rieu et al. 1989; Weliachew,
Casoli, \& Combes 1988) and 0.025 $\pm$ 0.003 at a resolution 
of $\sim 1\arcmin$ (1500 pc) (Helfer \& Blitz 1993). 
Figure \ref{n6946.ratio_vs_r}$a$ shows this radial
distribution of the average integrated I$_{\rm HCN}$/I$_{\rm CO}$ ratio.  
It is important to note that the points shown in Figure 
\ref{n6946.ratio_vs_r}$a$ 
represent the {\it average}  ratios over the area enclosed at the radius $r$,
i.e. that the plot represents the {\it integrated} ratios 
I$_{\rm HCN}$/I$_{\rm CO}$
as a function of $r$.  The monotonic falloff in the ratio with radius is 
simply a result of the confinement of the HCN emission to the inner $r$ = 150 
pc, while the CO is distributed over a much larger radius (there is detectable
emission at least to $r$ = 3.5 kpc, Tacconi \& Young 1989).  What is perhaps
a more interesting quantity physically is the {\it annular} ratio 
measured as a function of radius; that is, if I$_{\rm HCN}$/I$_{\rm CO}$ 
is $0.10 \pm 0.01$ measured as an average from $0 < r < 150$ pc, what
is the value of I$_{\rm HCN}$/I$_{\rm CO}$ from $150 < r < 800$ pc 
(where 800 pc is the radius of the NRAO beam in Helfer \& Blitz
1993)?  We can set an upper limit to I$_{\rm HCN}$/I$_{\rm CO}$ in this
annulus by comparing the BIMA data with the NRAO 12 m HCN flux.  
Since the BIMA measurement recovered $0.81 \pm 0.17$ of the 
single-dish HCN flux measured at the NRAO 12 m (Table 3), let us assume that
20\% of the single-dish HCN flux is distributed at radii larger than the 
interferometer was able to measure, yet within the half power beam area of 
the the NRAO 12 m --- that is, radii within the annulus $150 < r < 800$ pc.  
(This is a conservative estimate, since any ``missing'' large-scale flux could
also contribute to the flux at the central position.)  If we then measure
the flux in the Regan \& Vogel CO map from $150 < r < 800$ pc, we find that 
the ratio I$_{\rm HCN}$/I$_{\rm CO}$ in this annulus is at most 0.01.  
This ratio is an order of magnitude lower than that measured over 
the central $r = 150$ pc, as shown in Figure \ref{n6946.ratio_vs_r}$b$.

\subsection{ Nondetections }

We did not detect HCN emission from NGC 4321 (M100), NGC 4527, or NGC 4569.
While it is possible that the dense structure in these galaxies is so extended
that the interferometer resolves out the single-dish HCN emission (for
these sources, the flux of structures larger than $\sim$ 20\arcsec\ is
attenuated by $\ga$ 50\%), it is 
also likely that the observations simply were not sensitive enough to detect 
the HCN from these sources.  For those observations which suffered from an
intermittent phase lock (see $\S$ 2.1), the antenna that was flagged was 
one of the two that made up the shortest baseline pair;  thus the
calibration solution may not have been reliable and also the zero 
spacing problem may have been exacerbated for these observations.

\section{ Discussion }

\subsection{ The Radial Dependence of Dense Gas Ratios }

Spectroscopic studies of CS and HCN emission in normal external galaxies suggest
that most spiral galaxies contain an appreciable amount of gas at densities
of $\rm \ga 10^5 ~cm^{-3}$ in their centers (Mauersberger et al. 1989; Sage, 
Solomon, \& Shore 1990; Nguyen-Q-Rieu et al. 1992; Israel 1992; Helfer 
\& Blitz 1993).  The maps in Figure 1 show the distribution of that dense 
gas, namely, that it is confined to the central kiloparsec of 
the sources imaged.   This situation appears to be very similar to that
seen in the Milky Way, where widespread emission from the dense gas tracers
CS (Bally et al. 1987) and HCN (Jackson et al. 1996) is found only
within the central $\sim$ 500 pc diameter.    

What does the distribution of dense gas tell us about the physical conditions 
in the molecular gas as a function of its location in a galaxy?  To 
investigate the physical conditions rather than the total gas content, 
we normalize the HCN to that of the CO and consider the ratio 
$\rm I_{HCN}/I_{CO}$.  If the kinetic temperature of the gas responsible for
the cospatial HCN and CO emissions is about the same, 
then $\rm I_{HCN}/I_{CO}$ may be
considered as a qualitative indicator of the density or thermal pressure in 
the gas.  (Indeed, since the J = 1 state lies only 5.5 K above the
ground state for CO and 4.3 K above ground for HCN, even rather cold
gas has the energy to populate the J = 1 state for both molecules.
It is the molecular density that is more important in determining
the excitation.  See Helfer \& Blitz 1996.)
In NGC 6946 ($\S$ 3.2), $\rm I_{HCN}/I_{CO}$ is 0.11 $\pm$ 0.01 
averaged over the central $r < 150$ pc, whereas we deduce an upper limit 
of $\rm I_{HCN}/I_{CO} \le 0.01$ averaged in the annulus $150 < r < 800$ pc,
a region that includes the inner part of the disk.
A similar radial dependence of $\rm I_{HCN}/I_{CO}$ is seen in the unusual 
Seyfert/starburst hybrid galaxy NGC 1068, where the ratio approaches 0.6 in the 
central $r$ = 175 pc (Helfer \& Blitz 1995), and the ratio falls off
monotonically to about 0.1 at the large reservoir of molecular gas at about 
1 kiloparsec from the nucleus.  In the Milky Way, the ratio $\rm I_{HCN}/I_{CO}$
is about 0.081\footnote{See note $c$ to Table 4.} $\pm$ 0.004 averaged over the 
central $r = 315$ pc (Jackson et al. 1996);
between $3.5 < r < 7$ kpc in the plane of the Milky Way, the average ratio
is $\sim$ 0.026 $\pm$ 0.008, and in solar neighborhood GMCs, we measure 
$\rm I_{HCN}/I_{CO}$ ratios of 0.014 $\pm$ 0.020 when averaged over
$\sim$ 50 pc GMCs (Helfer 1995 -- and these are upper 
limits to the ratio averaged over several hundred parsecs).   These numbers
are summarized in Table 4.

What seems apparent from these comparisons is that $\rm I_{HCN}/I_{CO}$ is 
a strong function of galactocentric radius -- or more precisely, that there
is at least a bimodal distribution in the ratio in normal galaxies: 
the ratio at the center is substantially higher than elsewhere in the
galaxy.  Even though local effects
can dominate on scales of individual GMCs ($\S$ 3.2), the average ratio
of dense gas emission is highest at a galaxy's center and drops at
larger distances from the center.  Furthermore, the general agreement
between the ratios in the central $\sim$ 500 pc of NGC 6946 and the Milky 
Way suggest that the physical conditions in the centers of the two galaxies 
are similar.

\subsection{ GMCs in the High Pressure Environments of Galactic Bulges}

The measurement that the ratio $\rm I_{HCN}/I_{CO}$ is 5 to 10 times higher 
in the bulge regions of the Milky Way and NGC 6946 than in their disks suggests
that the physical conditions of the molecular gas in the bulge and disk 
regions are very different.  Furthermore, the presence of dense gas on size 
scales of $\sim$ 500 pc suggests that the internal pressure is very high in
the molecular gas.  Let us assume that the intrinsic line widths of the clouds 
are dominated by nonthermal, bulk motions as in local clouds, and that
their intrinsic linewidths are $\ge$ 1 km~s$^{-1}$.  Although we cannot
model the density accurately with the observation of a single transition
of HCN or CO, a simple LVG analysis of the line ratios suggests that
a line ratio of $\rm I_{HCN}/I_{CO}$ = 0.11 implies densities of
10$^{4.2-5.2}$ cm$^{-3}$ for gas at a kinetic temperature in the range 
T$_{\rm K}$ = 15 -- 70 K.  These densities are consistent with what
is measured in the Milky Way bulge molecular clouds, where
n(H$_2$) $\ga 10^4 \rm ~cm^{-3}$ (G\"usten 1989).
If we take the gas densities in NGC 6946 to be
$\rm 10^{4.6} ~cm^{-3}$ , then the typical kinetic pressure throughout
the molecular gas is $\rho v^2/k \sim 1 \times 10^7$ cm$^{-3}$~K.

Spergel \& Blitz (1992) considered the effects of the
extended, hot coronal gas in the bulge of the Milky Way on the thin molecular
layer that is embedded within it, and they argued that the pressure
in the center of the Galaxy is two to three orders of magnitude greater
than that in the solar neighborhood.  How does such an extraordinary 
difference in the environmental pressure affect the properties of molecular
clouds?  In disk GMCs, the external pressure of the ambient ISM ($\rm \sim
10^4 ~cm^{-3}$~K) is small compared with the pressure from the self-gravity
of a cloud ($\rm \sim 10^5 ~cm^{-3}$~K), and a source of ``local''
pressure (i.e., ongoing or recent star formation) is required to support  
any localized high-density clumps within the cloud.  In the high-pressure 
($\rm \sim 5 \times 10^6 ~cm^{-3}$~K, estimated for the Milky Way from the 
X-ray measurements of Yamauchi et al. 1990) environments of bulge molecular 
gas, on the other hand, the GMCs need not be self-gravitating -- in fact, 
only the most massive clouds {\it could} have the Jeans masses required to be 
self-gravitating.  In bulge clouds, then, it is the external pressure of the 
environment that dominates, and these high pressures can support the dense gas 
throughout the molecular component (regardless of whether there is any active 
star formation in the GMCs).
 
These arguments are easily extended to observations of external galaxies,
which typically also have X-ray emission associated with their bulges
(Fabbiano, Kim, \& Trinchieri 1992).  From our observations of HCN in
the bulges of external galaxies, it appears that the physical conditions
in molecular gas in the centers of galaxies are much more similar to 
each other than they are to local GMCs in the Milky Way.  

\subsection{ Positional Offsets Between the Dense Gas and Radio Continuum
		Distributions }

In the Milky Way, there is a pronounced offset of $\sim$ 80 pc in the position 
of the peak radio continuum emission (Sgr A*, $l = 0\arcdeg, b = 0\arcdeg$) and
the centroid of the dense gas (traced by CS and HCN) distribution ($l = 
0.6\arcdeg, b = 0\arcdeg$).  
There is also evidence for an offset of $\sim$ 100 pc in NGC 1068 between the
peaks of the HCN emission and the radio continuum emission (Helfer \& Blitz 
1995), and a kinematic analysis of the molecular gas suggests the existence of 
an $m = 1$ mode (i.e. a dipole asymmetry in the mass distribution) in NGC 1068.
Whether it is the radio continuum emission or the molecular gas that traces the
center of the mass distribution, the dynamical timescale of the offsets
is brief ($<$ 10$^6$ years).  If the offsets are indeed ephemeral, and
not a steady-state condition of the galaxies, then it is appropriate to look for 
a source of the instability that causes them.

How common are such offsets between the peaks of the dense molecular
gas distribution and the radio continuum emission in galaxies?
We compared the positions of peak HCN emission with those of the peak
radio continuum in our sample (Figure 1, Table 5).  While three of the galaxies
show a reasonable coincidence between the two positions, two of the five 
sources, NGC 4826 and M51, show significant offsets:  in NGC 4826,
the offset is $3\arcsec.7$ or 74 pc; in M51, it is $2\arcsec.6$ or 130 pc.   
(In M83, there is a $1\arcsec.6$ or 40 pc offset, but because of the elongated
beams and extended distribution in both HCN and the radio continuum, 
the peak positions are less certain.)

It appears that these offsets are a common feature in spiral galaxies.
In this sample, two out of five detected galaxies have offsets;  we have
already mentioned the offsets in the Milky Way and in NGC 1068.
In a recent study of the K-band morphology in 18 face-on spiral
galaxies, Rix \& Zaritsky (1995) found that about one
third of the disks have significant $m = 1$ modes at 2.5
disk exponential scale lengths.

It may be significant that the two galaxies in our sample that show
offsets, NGC 4826 and M51, also share the 
characteristic that they have low-level nonstellar nuclear activity 
(both are LINERs; these are low-level active galactic nuclei, or AGN); 
the other three galaxies detected here (NGC 3628, 
M83, and NGC 6946) have low-level starburst activity instead (L. Ho, 
private communication).    A recent study by Ho et al. (1993)
suggests that up to 80\% of the 500 brightest galaxies in the
northern sky harbor some kind of activity in their nuclei; of these,
about half are classified as LINERs, and half are starbursts.  It is
surprising both that {\it most} spiral galaxies appear to show
some kind of nuclear activity and also that the activity seems to
be roughly evenly divided between stellar and nonstellar mechanisms.
The offsets of the dense gas from the peak of the radio continuum
in NGC 4826 and M51, as well as that in NGC 1068 (a galaxy with a more
energetic AGN, but similar conceptually to the LINERs) may help to
distinguish empirically the kind of activity that dominates in a 
given galaxy's nucleus.

\section{ Conclusions }

We have presented the results of a survey of HCN emission from eight 
nearby spiral galaxies made with the BIMA interferometer.  These observations
comprise the first high-resolution survey of dense ($\sim 10^5$ cm$^{-3}$)
gas from a sample of relatively normal galaxies, rather than galaxies 
with prolific starburst or nuclear activities. 

We imaged five of the eight galaxies in HCN:  NGC 3628, NGC 4826, NGC 5194 
(M51), NGC 5236 (M83), and NGC 6946.  To within the uncertainties, the
interferometer recovers all of the single-dish flux measured for
each source in a single pointing at the NRAO 12 m telescope (Helfer \& 
Blitz 1993); this implies that there is not a significant contribution to 
the HCN fluxes from extended emission in the disks of the galaxies.
In all the galaxies observed, the HCN emission is confined to the central
kiloparsec of the sources.

We added zero-spacing data from the NRAO 12 m telescope to the BIMA CO map of 
NGC 6946 by Regan \& Vogel (1995) in order to compare the ratio of HCN to 
CO intensities in this galaxy.   The ratio $\rm I_{HCN}/I_{CO}$ ranges 
from 0.05--0.2 within the central $r = 150$ pc; the average ratio over this 
region is $\rm I_{HCN}/I_{CO} = 0.11 \pm 0.01$.  A comparison with single-dish
observations allows us to place an upper limit of $\rm I_{HCN}/I_{CO}
\le 0.01$ in the annulus $150 < r < 800$ pc in NGC 6946.

The extent of HCN emission in NGC 6946, $r \sim 150$ pc, and the ratio $\rm 
I_{HCN}/I_{CO} = 0.11$ in this region are similar to what is observed
in the Milky Way ($r \sim 250$ pc, $\rm I_{HCN}/I_{CO} = 0.08$, Jackson et al.
1996); this suggests that the physical conditions in the centers of these
two galaxies are similar.  Furthermore, in NGC 6946, NGC 1068, and the Milky 
Way, the ratios at the centers are 5 to 10 times higher than those in the disks
of these galaxies.  This result is consistent with an enhancement of two 
to three orders of magnitude in the pressure of the bulges compared with the 
disks (Spergel \& Blitz 1992).  It appears that the 
physical regions in the centers of galaxies are much more like each
other than the conditions in the center of a galaxy relative to its disk.  

In NGC 4826 and M51, as in the Milky Way and in NGC 1068, there is 
a linear offset of $\sim$ 100 pc between the dense gas distribution 
and the peak of the radio continuum emission.  These offsets appear
to be a common feature in galaxies and may indicate that their disks 
are non-axisymmetric. It may be significant that of the five
galaxies imaged in HCN, NGC 4826 and M51 are LINERs, whereas the other
three are starbursts.

We did not detect HCN in three galaxies with positive single-dish HCN emission:
NGC 4321 (M100), NGC 4527, and NGC 4569.   It could be
that the emission is extended enough in these sources that the interferometer
resolved out any detectable emission; however, we cannot rule out the
possibility that there was some intrinsic problem with the observations
of these sources.

\acknowledgments
We thank the referee, Paul Ho, for his careful reading and suggestions;
these helped us to improve the manuscript.
We thank Mike Regan for providing us with the BIMA NGC 6946 CO data, 
and we thank Darrel Emerson, Phil Jewell, 
Tom Folkers and the staff of the NRAO 12 m telescope for assistance 
with the OTF observations and data reduction.  Luis Ho helped with
the early stages of the BIMA observations.  We thank Kotaro Kohno
for kindly providing us with the NRO map of HCN in M51 prior to publication.  
TTH thanks Jack Welch for hospitality while visiting UC-Berkeley.  This 
research was partially supported by a grant from the National Science 
Foundation, with additional support from the State of Maryland.

\clearpage

%%*********************************************************************
%%
%% Table 1: Details of the observations, BIMA HCN spiral galaxies 
%%
%%*********************************************************************

\begin{deluxetable}{lrrrr}
%%\tablewidth{42pc}
\tablewidth{5.8truein}
\tablecaption{Sources }
\tablehead{
\colhead{Source}                &
\colhead{$\alpha$(J2000)}       &
\colhead{$\delta$(J2000)}       &
\colhead{$d$}			&
\colhead{$v$$_{\rm LSR}$\tablenotemark{a}}      \\[.2ex]
\colhead{}			&
\colhead{}			&
\colhead{}			&
\colhead{(Mpc)}			&
\colhead{(km~s$^{-1}$)}		
}
 
\startdata
NGC 3628\tablenotemark{b}       & 11$^{\rm h}$20$^{\rm m}$16.$^{\rm s}$27 & 
	13$^{\arcdeg}$35$^{\arcmin}$39.$^{\arcsec}$0  & 9 & 847  \nl
NGC 4321 (M100)& 12 22 54.80 & 15 49 23.0  & 17  & 1550 \nl
NGC 4527       & 12 34 08.80 & 02 39 10.3  & 20  & 1734 \nl
NGC 4569       & 12 36 50.02 & 13 09 53.1  & 17  & -235 \nl
NGC 4826       & 12 56 44.25 & 21 40 52.3  & 5   & 408  \nl
NGC 5194 (M51) & 13 29 53.30 & 47 11 50.0  & 10  & 463  \nl
NGC 5236 (M83) & 13 37 00.23 & -29 52 04.5 & 5   & 516  \nl
NGC 6946       & 20 34 51.91 & 60 09 11.9  & 5   & 52   \nl
\enddata
\tablenotetext{a}{ $v$$_{\rm LSR}$ is defined by the radio convention, 
$v_{\rm LSR}$/c = $\Delta$$\lambda$/$\lambda$$_o$, where $\lambda$$_o$ 
is the wavelength in the rest frame of the source.}
\tablenotetext{b}{ Pointing center is the optical center of NGC 3628.
The radio continuum center is some 21\arcsec\ to the SE of the optical
center and is coincident with the HCN emission (see $\S$ 3.1).}
\end{deluxetable}

%%*********************************************************************
%%
%% Table 2: Map parameters of BIMA HCN spiral galaxy observations
%%
%%*********************************************************************

\begin{planotable}{lrrr}
%%\tablewidth{42pc}
\tablewidth{5.8truein}
\tablecaption{Mapping Details}
\tablehead{
\colhead{Source}                &
\colhead{Beam}                  &
\colhead{K Jy$^{-1}$}           &
\colhead{$\sigma$$_{\rm mom}$\tablenotemark{a}}   \\[.2ex]
\colhead{}	&
\colhead{($\arcsec$ $\times$ $\arcsec$)} &
\colhead{} &
\colhead{(Jy bm$^{-1}$ km s$^{-1}$)}  
}
 
\startdata
NGC 3628       & 4.7  $\times$ 4.1  & 8.13 & 2.0  \nl
NGC 4321 (M100)& 8.2  $\times$ 6.5  & 2.91 & 2.1  \nl
NGC 4527       & 13.5 $\times$ 10.7 & 1.07 & 3.6  \nl
NGC 4569       & 5.5  $\times$ 4.2  & 6.85 & 1.0  \nl
NGC 4826       & 10.5 $\times$ 8.0  & 1.87 & 3.2  \nl
NGC 5194 (M51) & 7.9  $\times$ 6.5  & 3.03 & 1.5  \nl
NGC 5236 (M83) & 12.5 $\times$ 4.1  & 3.03 & 3.6  \nl
NGC 6946       & 5.9  $\times$ 5.0  & 5.31 & 1.7  \nl
\enddata
\tablenotetext{a}{$\sigma$$_{\rm mom}$ is lower than the statistical
noise in the integrated intensity (``moment'') maps because emission 
below the level of the statistical noise in the channel maps was masked 
out in order to calculate the moment maps (see text).} 
\end{planotable}

%%*********************************************************************
%%
%% Table 3: Comparison of BIMA HCN spiral galaxy observations with single dish
%%
%%*********************************************************************

\begin{planotable}{lrrr}
%%\tablewidth{42pc}
\tablewidth{5.8truein}
\tablecaption{Comparison with Single Dish Observations}
\tablehead{
\colhead{Source}                &
\colhead{I$_{\rm BIMA}$\tablenotemark{a}}        &
\colhead{I$_{\rm KP}$\tablenotemark{b}}  &
\colhead{Fraction\tablenotemark{c}}   	          \\[.2ex]
\colhead{}                      &
\colhead{(Jy~kms$^{-1}$)}         &
\colhead{(K~kms$^{-1}$)}         &
\colhead{recovered} 
}

\startdata
NGC 3628       & 49.0 $\pm$ 17.5 & 1.5 $\pm$ 0.2 & 1.01 $\pm$ 0.36 \nl
NGC 4826       & 39.0 $\pm$ 16.7 & 1.3 $\pm$ 0.2 & 0.93 $\pm$ 0.41 \nl
NGC 5194 (M51) & 26.4 $\pm$ 10.5 & 1.2 $\pm$ 0.2 & 0.68 $\pm$ 0.27 \nl
NGC 5236 (M83) & 103  $\pm$ 23   & 2.7 $\pm$ 0.2 & 1.18 $\pm$ 0.27 \nl
NGC 6946       & 39.2 $\pm$ 5.5  & 1.5 $\pm$ 0.2 & 0.81 $\pm$ 0.17 \nl
\enddata
\tablenotetext{a}{ I$_{\rm BIMA}$ from this study, corrected for primary
beam effects from the BIMA interferometer and the NRAO 12 m telescope for 
comparison with I$_{\rm KP}$.}
\tablenotetext{b}{ I$_{\rm KP}$ from Helfer \& Blitz 1993, corrected to main 
beam brightness temperature scale.}
\tablenotetext{c}{ Fraction recovered = (I$_{\rm BIMA}$ $\times$ K~Jy$^{-1}$ 
$\times$ $\theta$$_1$$\theta$$_2$/(71\arcsec)$^2$) / I$_{\rm KP}$, where
$\theta$$_1$, $\theta$$_2$, and K~Jy$^{-1}$ are listed in Table 2.}
\end{planotable}

%%*********************************************************************
%%
%% Table 4: HCN/CO ratios in MW, NGC6946 and NGC 1068 
%%
%%*********************************************************************

\begin{planotable}{lccc}
%%\tablewidth{42pc}
\tablewidth{5.8truein}
\tablecaption{Differential I$_{\rm HCN}$/I$_{\rm CO}$ Ratios} 
\tablehead{
\colhead{Region}                &
\colhead{Milky Way} 		&
\colhead{NGC 6946}              &
\colhead{NGC 1068\tablenotemark{a}}              
}
\startdata
Bulge{\tablenotemark{b}}& 0.081{\tablenotemark{c}} $\pm$ 0.004 & 0.11 $\pm$ 0.01  & 0.6 \nl
Disk{\tablenotemark{d}} & 0.026{\tablenotemark{e}} $\pm$ 0.008 & $\le$ 0.01 & 0.1 \nl
Local GMCs & 0.014 $\pm$ 0.020{\tablenotemark{e}} & ---        & --- \nl
\enddata
\tablenotetext{a}{Helfer \& Blitz 1995}
\tablenotetext{b}{Milky Way:  r = 300 pc; NGC 6946: r = 150 pc; NGC 1068: r =
175 pc}
\tablenotetext{c}{Jackson et al. 1995.  We use their correction for HCN
and CO emission at nonzero Galactic latitude.  In their ``aperture 
photometry'', Jackson et al. smooth their HCN map to a spatial
resolution larger than that of their CO map by a factor of 
$\rm \lambda(HCN)/\lambda(CO)$ in order to emulate the single-dish 
measurements of extragalactic ratios.  This correction is not appropriate
for comparison with our interferometeric measurements, in which the
HCN and CO maps have been smoothed to the same resolution.  We therefore
``re-correct'' their ratio so that it is appropriate for direct comparison to
these results. }
\tablenotetext{d}{Milky Way: $3.5 < r < 7$ kpc; NGC 6946: $150 < r < 800$ pc;
NGC 1068: $1.0 < r < 1.4$ kpc}
\tablenotetext{e}{Helfer 1995}
\end{planotable}

%%*********************************************************************
%%
%% Table 5: Positional offsets between BIMA HCN and radio continuum positions 
%%
%%*********************************************************************

\begin{planotable}{lr}
%%\tablewidth{42pc}
\tablewidth{3.8truein}
\tablecaption{Offsets Between HCN and Radio Continuum Emission} 
\tablehead{
\colhead{Source}                &
\colhead{$\Delta\alpha,\Delta\delta$} \\[.2ex]
\colhead{}                      &
\colhead{($\arcsec \times \arcsec$)}         
}
\startdata
NGC 3628       & -0.5, -0.1 \nl
NGC 4826       & +2.2, -3.0 \nl
NGC 5194 (M51) & -2.2, +1.3 \nl
NGC 5236 (M83) & -1.3, -0.9 \nl
NGC 6946       & -0.1, -0.5 \nl
\enddata
\tablenotetext{}{References for Radio Continuum Data: NGC 3628, NGC 5236 and
NGC 6946: Condon et al. 1982; NGC 4826: Braun et al. 1994; NGC 5194: Turner
\& Ho 1994  }
\end{planotable}

\clearpage

\begin{figure}
\vspace{-1.5in}
\plotfiddle{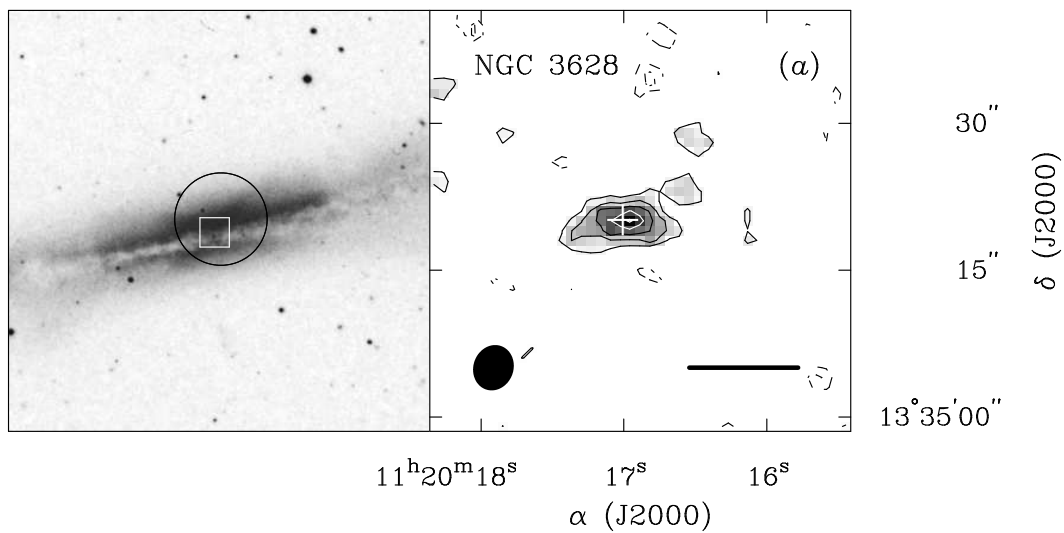}{300pt}{-90}{60}{60}{-230}{300}
\caption{BIMA images of HCN in ($a$) NGC 3628, ($b$) NGC 4826,
($c$) NGC 5194 (M51), ($d$) NGC 5236 (M83), and ($e$) NGC 6946.  The left 
panel in each plot
shows a 10\arcmin\ $\times$ 10\arcmin\ field from the Digitized Sky Survey
around the center of each galaxy. The circle represents the BIMA primary 
beam size of 132\arcsec\ FWHM, and the white square shows the field 
presented in the right panel.  The right panel shows the BIMA HCN image of 
each source.  The FWHM size of the synthesized beam is shown in the lower 
left corner of each image, and the horizontal bar represents a linear size 
scale of 500 pc.  The contour levels are $\pm$ 2,3,4... $\sigma_{\rm mom}$, 
where $\sigma_{\rm mom}$ is listed in Table 2 for each source.   The white
cross marks the position of the peak radio continuum emission (see text).
The images have not been corrected for primary beam attenuation. }
\label{hcnmaps}\end{figure}

\begin{figure}
\figurenum{1}
\vspace{-1.5in}
\plotfiddle{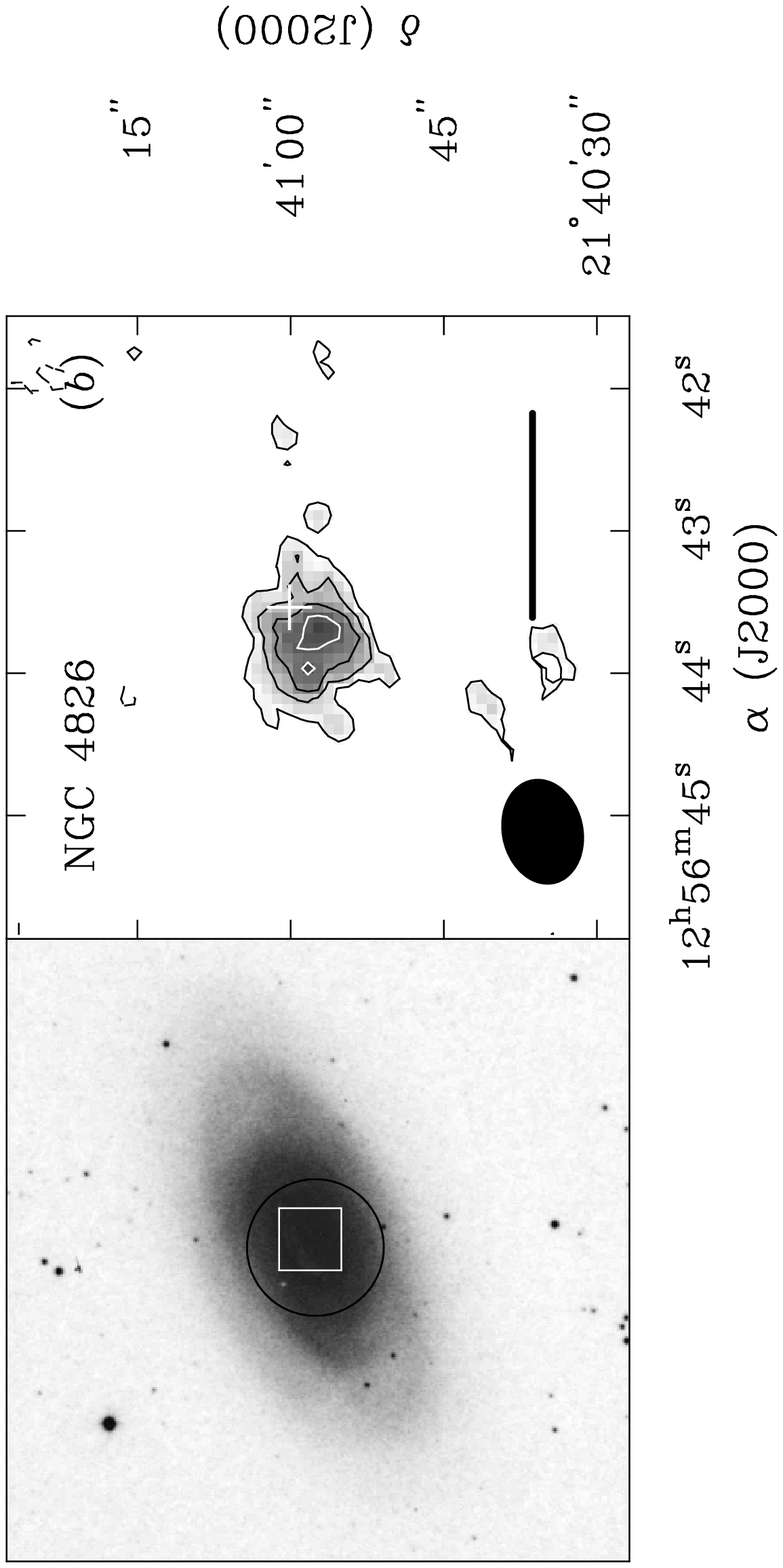}{300pt}{-90}{60}{60}{-230}{300}
\plotfiddle{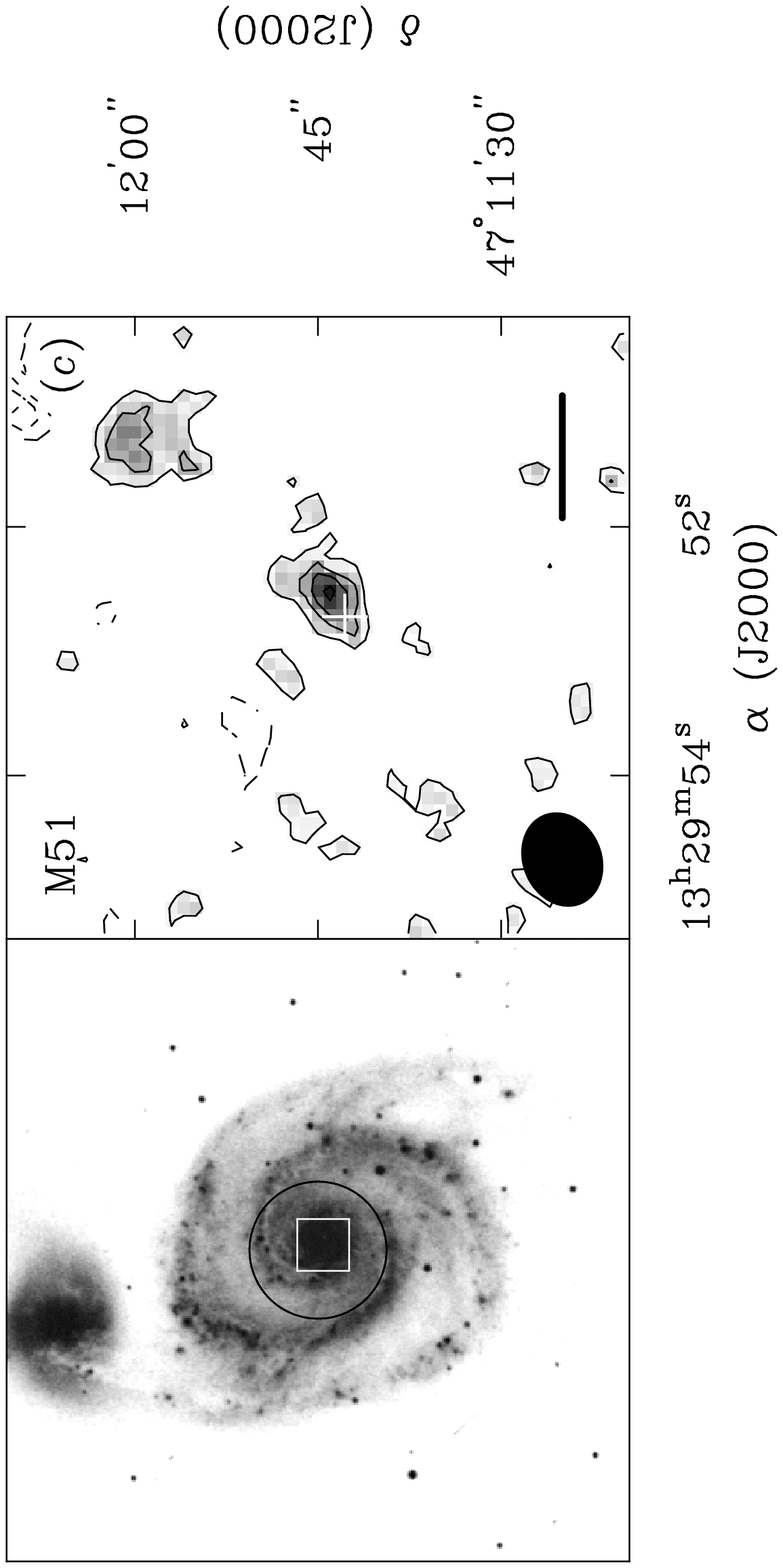}{300pt}{-90}{60}{60}{-230}{370}
\caption{continued}
\end{figure}
 
\begin{figure}
\figurenum{1}
\vspace{-1.5in}
\plotfiddle{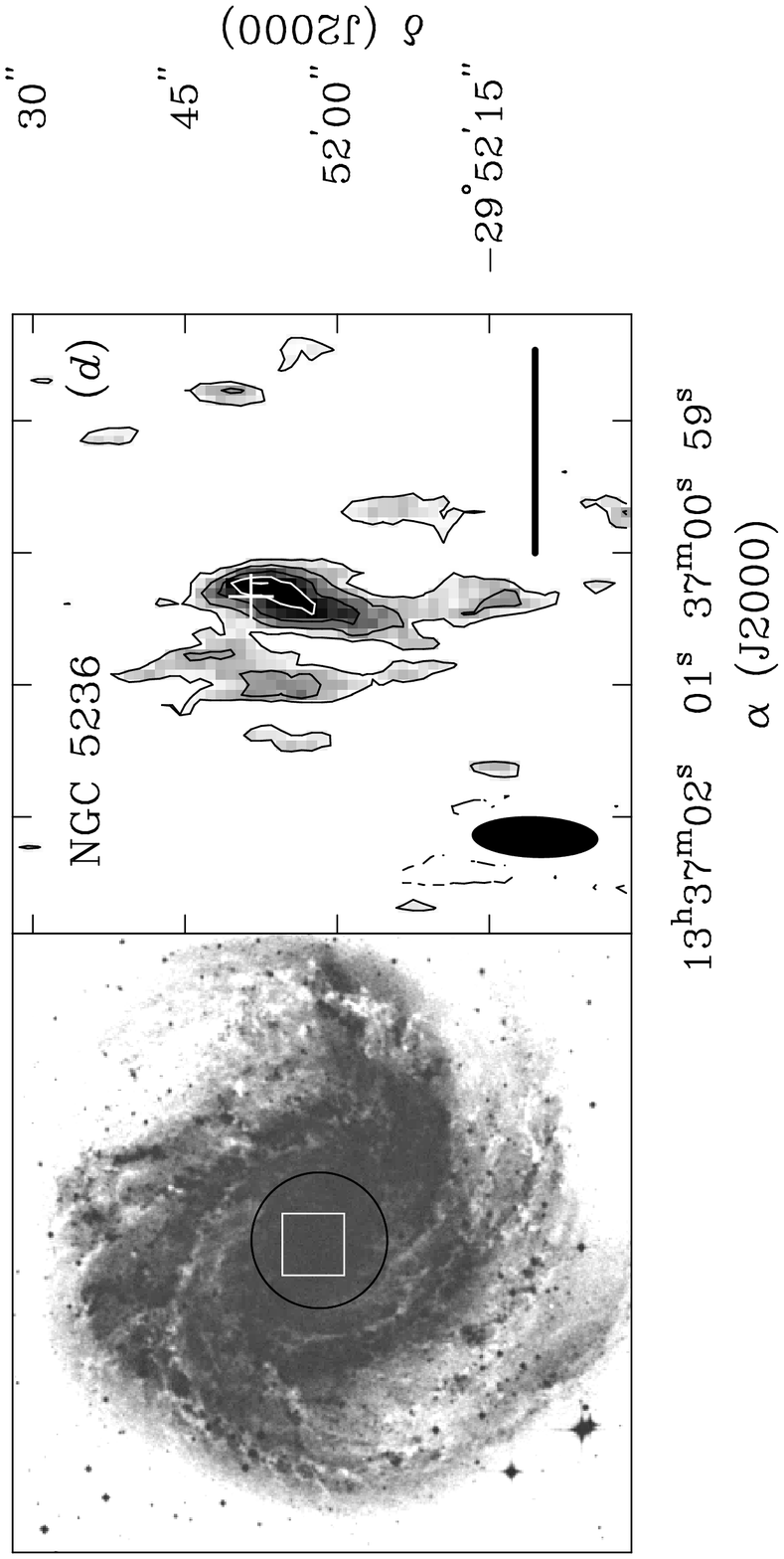}{300pt}{-90}{60}{60}{-230}{300}
\plotfiddle{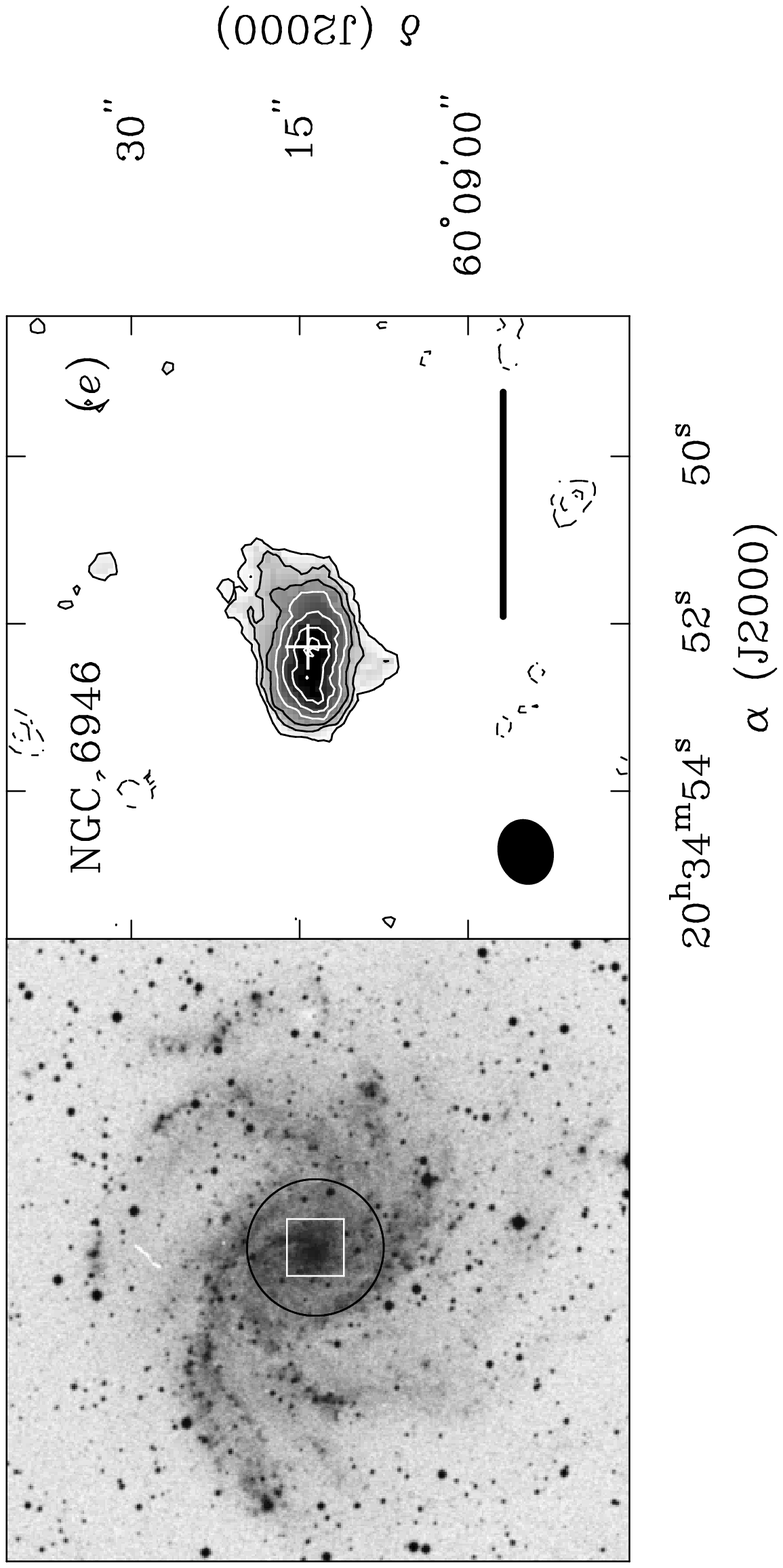}{300pt}{-90}{60}{60}{-230}{370}
\caption{continued}
\end{figure}

\begin{figure}
\plotfiddle{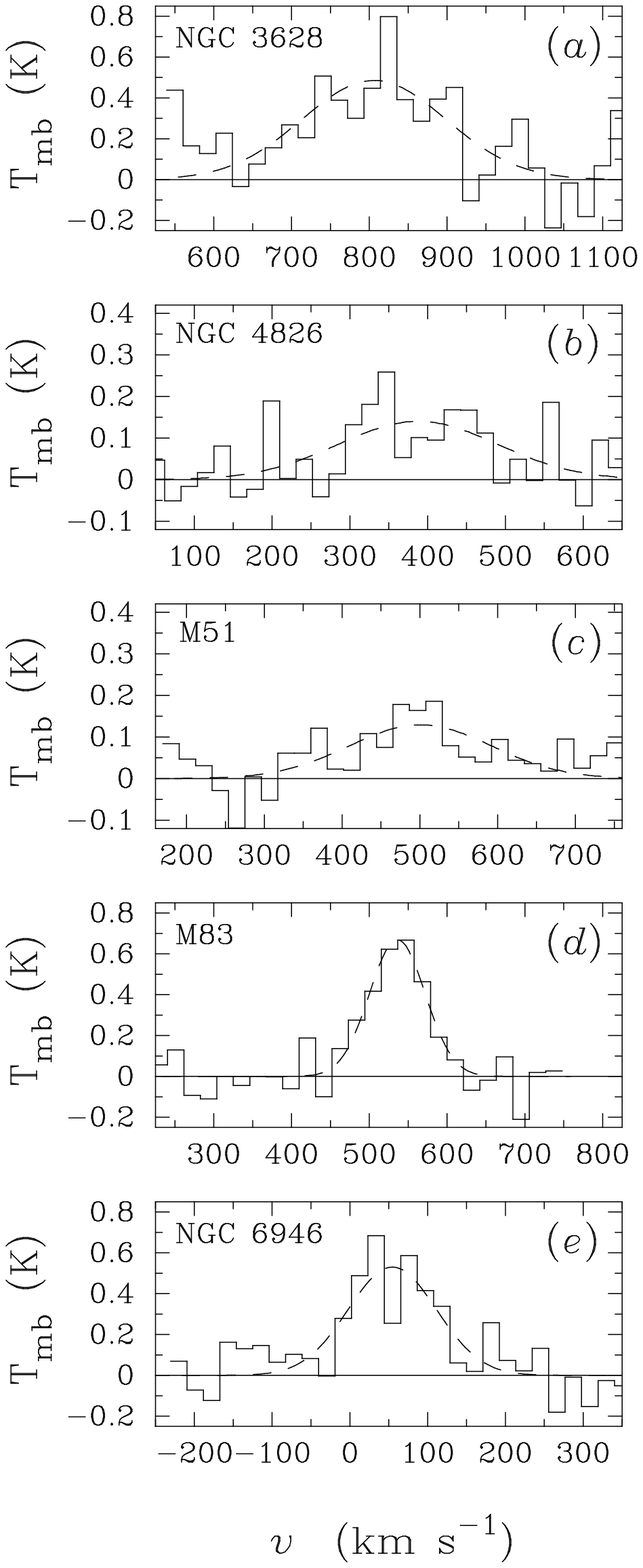}{400pt}{0}{80}{70}{-170}{-30}
\caption{Spectra from the positions of peak HCN emission in 
($a$) NGC 3628, ($b$) NGC 4826, ($c$) NGC 5194 (M51), ($d$) NGC 5236 (M83), 
and ($e$) NGC 6946.  The abscissa is the radio-defined LSR velocity in 
km s$^{-1}$, and the ordinate is main beam brightness temperature in K. }
\label{spectra}\end{figure}

\begin{figure}
\vspace{-1.5in}
\plotfiddle{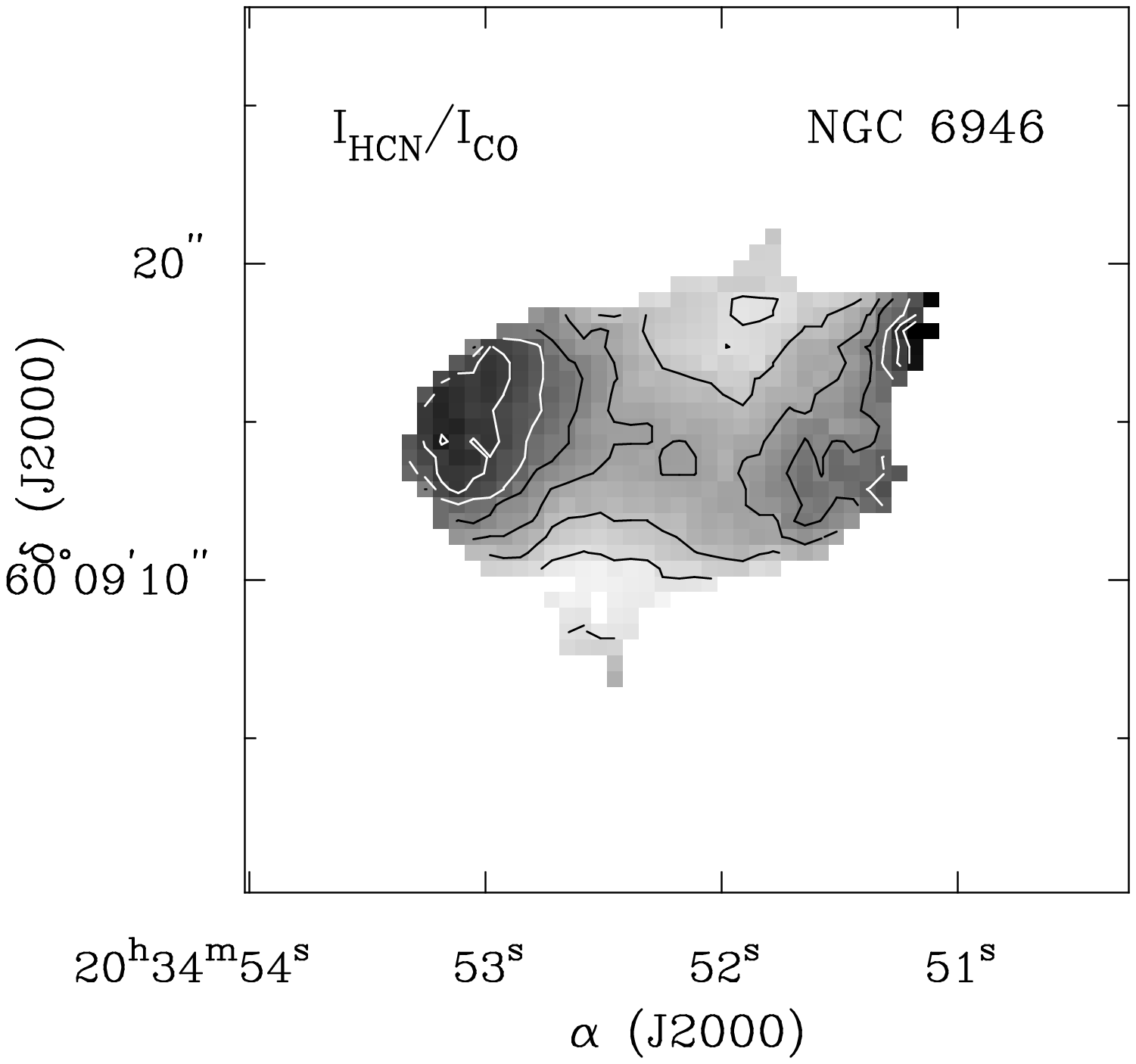}{400pt}{0}{70}{70}{-200}{-30}
\caption{Ratio of HCN/CO integrated intensities in the
central $r = 150$ pc of NGC 6946.  The halftone limits are (0.04,0.2). 
The lowest contour level is 0.06, and the contour interval is 0.02.  
The CO map used to construct the ratio map includes zero-spacing data 
from the NRAO 12 m telescope.}
\label{n6946rat}\end{figure}

\begin{figure}
\plotfiddle{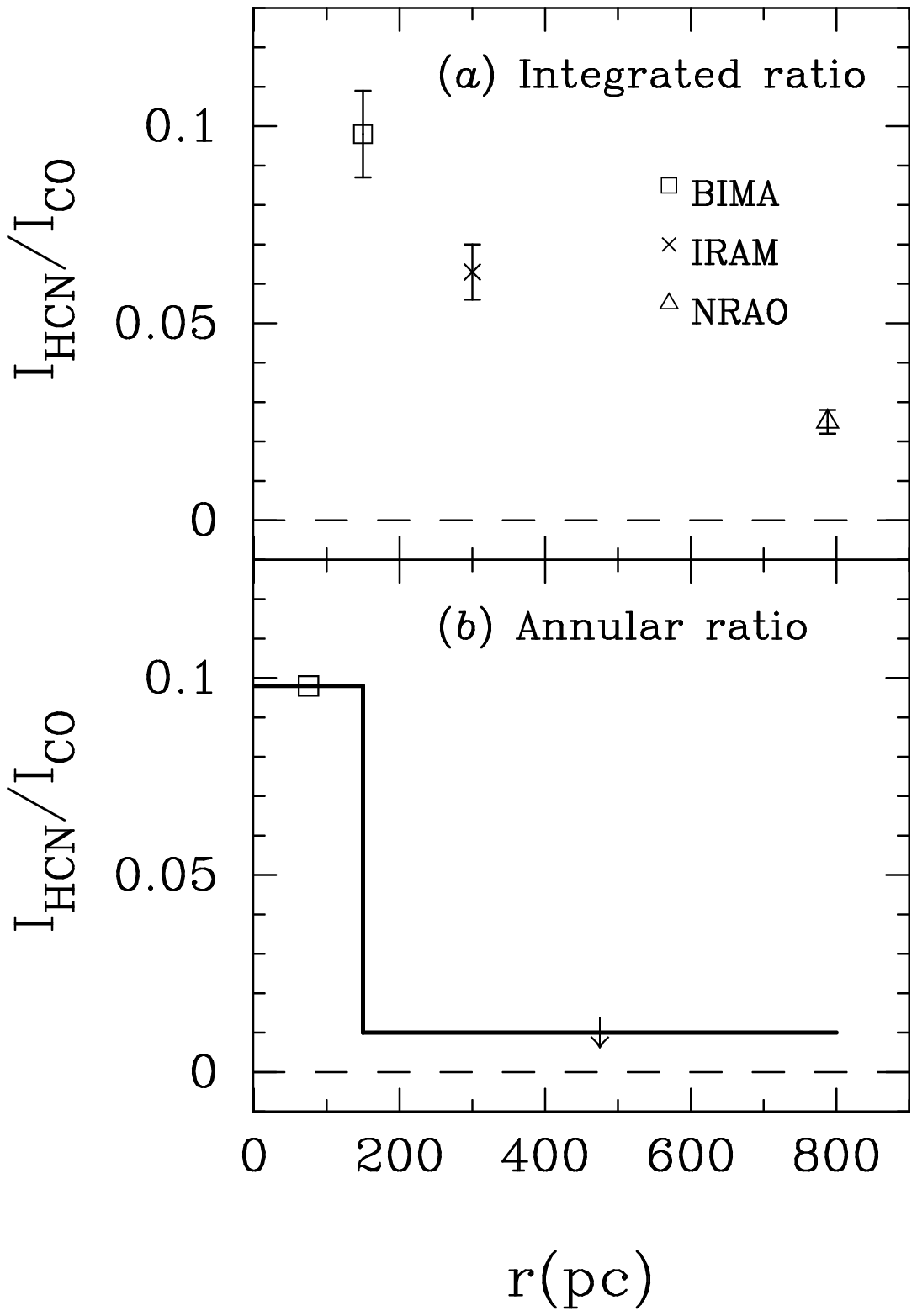}{400pt}{0}{70}{70}{-230}{-70}
\caption{ ($a$) The integral ratio of HCN/CO integrated intensities as a
function of radius in NGC 6946.  The innermost point is the BIMA measurement
from this study, the point at $r$ = 300 pc uses data from the IRAM 30 m
telescope (Nguyen-Q-Rieu et al. 1992; Weliachew et al. 1988), and the
outermost point was measured using the NRAO 12 m telescope (Helfer \& 
Blitz 1993). Each point represents the average ratio over the area enclosed 
at the shown radius. ($b$) The more physical differential or annular ratio 
as a function of radius in NGC 6946.  The figure emphasizes the sharp 
boundary between the physical conditions
in the $r$ = 300 pc HCN-emitting region and those at larger radii. 
}
\label{n6946.ratio_vs_r}\end{figure}

\end{document}